\documentclass{article}

\usepackage{graphics}
\usepackage{graphicx}
\usepackage{epsfig}
\usepackage{subfigure}
\usepackage{graphics}
\usepackage{amsmath,amsfonts,amssymb}
\usepackage{multirow}
\everymath{\displaystyle}

\title{A Sensitivity Matrix Based Methodology for Inverse Problem Formulation}

\author{ Ariel Cintr\'{o}n-Arias$^{1,2}$, H. T. Banks$^{1,2}$, Alex Capaldi$^{1}$, Alun L. Lloyd$^{1,2}$ \\ \\
$^{1}$Center for Research in Scientific Computation \\
North Carolina State University, Raleigh, NC 27695\\ \\
$^{2}$Center for Quantitative Sciences in Biomedicine\\
North Carolina State University, Raleigh, NC 27695\\ \\
}

\begin{document}
\date{ April 10, 2009}
\maketitle{}

\begin{abstract}We propose an algorithm to select parameter subset combinations
that can be estimated using an ordinary least-squares (OLS)
inverse problem formulation with a given data set. First, the
algorithm selects the parameter combinations that correspond to
sensitivity matrices with full rank. Second, the algorithm
involves uncertainty quantification by using the inverse of the
Fisher Information Matrix.  Nominal values of parameters are used
to construct synthetic data sets, and explore the effects of
removing certain parameters from those to be estimated using OLS
procedures. We quantify these effects in a score for a vector
parameter defined using the norm of the vector of standard errors
for components of estimates divided by the estimates. In some
cases the method leads to reduction of the standard error for a
parameter to less than 1\% of the estimate.
\end{abstract}



Acknowledgments: A.C.-A. is grateful to Dr. Sava Dediu
for helpful discussions about sensitivity identifiability that led
to the initial version of this manuscript.  This research was
supported in part by Grant Number R01AI071915-07 from the National
Institute of Allergy and Infectious Diseases and in part by the
Air Force Office of Scientific Research under grant  number
FA9550-09-1-0226. The content is solely the responsibility of the
authors and does not necessarily represent the official views of
the NIAID, the NIH or the AFOSR.

%

\clearpage
\section{Introduction}

The question of parameter identifiability/estimation in the
context of parameter determination from system observations or
output is at least forty years old and received much attention in
the zenith years of linear system and control theory in the
investigation of observability, controllability and detectability
\cite{AE,bellams,BeKa,Eykhoff,GW,Kalman,MehraLain,reid77,SageMelsa}.
These early investigations and results were focused primarily on
engineering applications, but much interest in other areas (e.g,
oceanography, biology) has prompted more recent inquiries for both
linear and nonlinear dynamical systems
\cite{anh06,BS,cob80,evans05,holm82,navon,white01,wu08,xia03,yue08}.
In some of the earliest results, Bellman and Astrom \cite{bellams}
defined the concept of structural identifiability, and provided a
theoretical framework to address, a priori, whether or not it is
possible to determine estimates of unknown parameters from
experimental data. Specifically they showed that controllablility
(in the sense of the Kalman \cite{Kalman} controllability matrix
possessing full rank) implies identifiability, thereby
establishing one of the earliest linear algebraic tests for
identifiability. In another important early linear algebraic
effort, Reid \cite{reid77} defined the term {\em sensitivity
identifiability} . If $z(\theta)$ denotes the output of a model
depending on a parameter vector $\theta$, then Reid explains {\em
sensitivity identifiability} in the following way.  Let $\Delta
\theta$ denote a local perturbation about a nominal $\theta_0$,
i.e., $\Delta \theta=\theta-\theta_0$, which gives rise to local
(small) perturbation $\Delta z$ in the output, i.e., $\Delta
z=z(\theta)-z(\theta_0)$.  Suppose that $\chi=\frac{\partial
z}{\partial \theta}$ denotes the sensitivity matrix, i.e., the
Jacobian matrix of the output, being evaluated at $\theta_0$
\cite{banksdav, banksded}. Then the first order Taylor
approximation (exact for linear dependence on the parameter)
\begin{equation}\label{seniden}
        \Delta z \approx \chi \Delta\theta.
\end{equation}
relates the perturbations.  A parameter vector is defined as {\em
sensitivity identifiable} if equation (\ref{seniden}) can be
solved uniquely (in the local sense) for $\Delta\theta$
\cite{cob80,reid77}. In their review, Cobelli and DiStefano
\cite{cob80} explain that a sufficient condition for sensitivity
identifiability is the nonsingularity of the matrix $\chi^T\chi$
or equivalently
 \[
        \det(\chi^T\chi)\neq0.
 \]
From this one sees immediately that parameter estimation depends
inherently on the condition number of the {\em Fisher Information
Matrix} (FIM) $F=\chi^T\chi$. Not surprisingly, subsequent
investigations of parameter estimation (in applied mathematics,
engineering, and statistics) have focused on the role of the FIM.
It is now well known that this matrix and its condition number
play a fundamental role in a range of useful ideas such as model
comparison \cite{BA} (the Akaike Information Criteria, the
Takeuchi Information Criteria, etc.), generalized sensitivity
functions \cite{banksded,BEG,TC} and experimental design
(duration, frequency, quality, etc., of observations required to
reliably estimate parameters) as well as computation of standard
errors and confidence intervals \cite{banksdav,BDE,BEG,DG}.

Brun, et al., \cite{brun} and Burth, et al., \cite{burth} proposed
analyses that use submatrices of the FIM $\chi^T\chi$. Burth, et
al., implement a {\em reduced-order} estimation by determining
which parameter axes lie closest to the ill-conditioned directions
of $\chi^T\chi$, and then by fixing the associated parameter
values at prior estimates throughout an iterative estimation
process.  Brun, et al., determined identifiability of parameter
combinations using the eigenvalues of submatrices that result from
only using some columns of $\chi^T\chi$. Motivated by these
efforts and those on the relationship between ill-conditioning of
the FIM and quality of parameter estimates investigated in
\cite{BDE,banksded,BEG}, we here use the sensitivity matrix $\chi$
to develop a methodology to assist one in parameter estimation or
inverse problem formulations.

In particular, in this paper we investigate the problem of finding
multiple solutions for unknown parameters from observations with a
statistical error structure (a more practical setting than one
assuming noise free observations). We address parameter
identifiability by exploiting properties of {\em both} the
sensitivity matrix and uncertainty quantifications in the form of
standard errors. We propose an algorithm inspired by
\cite{brun,burth}, to select parameter combinations (vectors) in
two stages.  In the first stage, all possible parameter
combinations (i.e., subsets of all parameters) are considered and
only those with a full rank sensitivity matrix are selected. In
the second stage, a score involving uncertainty quantification
(standard errors) is calculated for each parameter vector selected
in the first stage. Then parameter subset combinations are
examined in view of their score and the condition number of
corresponding sensitivity matrices. We believe that some form of
this type of {\em practical identifiability analysis} could be
carried out a priori, i.e., before any attempt to solve inverse
problems (from experimental observations) is made. We illustrate
the ideas and methodology with a seasonal epidemic model.

This manuscript is organized in the following manner. Section
\ref{seirsmdl} introduces the seasonal epidemic model. In Section
\ref{statmdl} we explain the statistical model for the observation
process; we define the ordinary least squares (OLS) estimator and
define precisely the Fisher information matrix. Using a first
order Taylor expansion of the model output we compute the OLS
estimator in terms of the sensitivity matrix singular values and
the error in the observation process. Section \ref{selalgo}
contains the proposed subset selection algorithm. In Section
\ref{appl} some illustrations of the algorithm are discussed, in
light of using both synthetic and observational data sets. We
conclude with a brief discussion in the last section.

\section{Motivating seasonal SEIRS model with demography}\label{seirsmdl}

We introduce a specific model, a standard {\em
Susceptible-Exposed-Infective-Recovered-Susceptible} (SEIRS)
model, to illustrate the methodology we discuss in this paper. In
particular we consider a seasonal model for disease spread and
progression in a population. Seasonal patterns of disease
incidence are observed in epidemics of influenza \cite{dusplo},
meningococcal meningitis \cite{riepil}, measles \cite{andgren},
and rubella \cite{witkar}, to mention a few.  Many temporal
factors play a role in the formation of cyclical patterns, for
instance \cite{grasfras}: (i) survival of the pathogen outside the
host, (ii) host behavior and (iii) host immune function.

Cyclical incidence patterns are often modeled with a transmission
parameter being a function of time.  We denote the time-dependent
transmission parameter by $\beta(t)$; it is traditionally defined
by \cite{dusplo,kalisip}
\begin{equation}\label{betdet}
 \beta(t)=\beta_0\left[1+\beta_1\cos(2\pi(t-t_0))\right],
 \end{equation}
where $\beta_0$ is called the baseline level of transmission,
$\beta_1$ is known as the amplitude of seasonal variation or
simply the strength of seasonality, and $t_0$ denotes the
transmission parameter phase shift.  We may, for convenience,
derive an equivalent formulation. Because
 \[
  \beta_1\cos(2\pi(t-t_0))=a_1\cos(2\pi t)+b_1\sin(2\pi t),
 \]
where $a_1=\beta_1\cos(2\pi t_0)$ and $b_1=\beta_1\sin(2\pi t_0)$,
we may re-write equation (\ref{betdet}) as
\begin{equation}\label{betatredef}
\beta(t)=\beta_0\left(1+a_1\cos(2\pi t)+b_1\sin(2\pi t)\right).
\end{equation}

The time-dependent transmission parameter $\beta(t)$, as defined
in equation (\ref{betatredef}), is used in the seasonal  epidemic
model introduced here.  Four main epidemiological events are
described: latent infection, active infection, recovery, and loss
of immunity.  It is assumed that individuals becoming infected
undergo latency, a period of time during which they are incapable
of effectively transmitting the infectious agent, before
progressing into active infection. People recover from active
infection and develop temporary immunity (they will eventually
become susceptible once again). Four epidemiological classes are
considered, and at time $t$ the number of: susceptible is denoted
by $S(t)$; latent or exposed is denoted by $E(t)$; infectious is
denoted by $I(t)$; and recovered or temporarily immune is denoted
by $R(t)$.  The nonlinear differential equations
\cite{kuzpic,schsmi}
\begin{eqnarray}
    \label{seqn}
    \frac{dS}{dt}&=& \frac{1}{P}N +\frac{1}{L}R(t)-\beta(t) S(t)\frac{I(t)}{N} -\frac{1}{P}S(t)\\
    \label{eeqn}
    \frac{dE}{dt}&=&\beta(t) S(t)\frac{I(t)}{N}-\frac{1}{M}E(t) -\frac{1}{P} E(t)\\
    \label{ieqn}
    \frac{dI}{dt}&=& \frac{1}{M}E(t)-\frac{1}{D}I(t) -\frac{1}{P} I(t)\\
    \label{xeqn}
    \frac{dR}{dt}&=&\frac{1}{D}I(t)-\frac{1}{L}R(t)-\frac{1}{P} R(t)\\
    N&=&S(t)+E(t)+I(t)+R(t)\\
    S(t_0)&=&S_0\\
    E(t_0)&=&E_0\\
    I(t_0)&=&I_0\\
    R(t_0)&=&N-S_0-E_0-I_0,
\end{eqnarray}
define the epidemic dynamics known as an SEIRS model.  This
formulation takes into account demographic processes (the birth
rate is $N/P$ and the average life span is $P$) while assuming the
total population size $N$ remains constant.

The mean latency period is denoted by $M$, while the average length
of active infection is denoted by $D$.  It is also assumed immunity lasts an average of
$L$ units of time.

In this paper we consider a scenario where the initial conditions
of the SEIRS model ($S_0$, $E_0$, and $I_0$) may be unknown, and
may need to be estimated, along with all the other model
parameters.  We apply inverse problem methodologies to determine
estimates of the vector parameter
\begin{equation}
    \theta=(S_0,E_0,I_0,N,L,D,M,P,\beta_0,a_1,b_1)^T\in\mathbb{R}^{p}=\mathbb{R}^{11},
\end{equation}
according to an ordinary least squares criterion (defined in the next section).

\section{Statistical model for the observation process}\label{statmdl}

The observation process is formulated assuming the SEIRS model,
together with a particular choice of parameters (the \lq\lq true"
parameter vector denoted as $\theta_0$) describes the epidemic
process exactly, but that the $n$ longitudinal observations
$\left\{Y_j\right\}_{j=1}^n$ are affected by random deviations
(such as measurement errors) from this underlying process.  More
precisely, if $z(t_j;\theta_0)$ denotes the number of new cases of
active infection (also referred to as the model output) between
the observation time points $t_{j-1}$ and $t_j$, which is defined
as
\begin{equation}
z(t_j;\theta_0)=\int_{t_{j-1}}^{t_j} \frac{1}{M} E(t;\theta_0)dt,
\end{equation}
then the statistical model for the observation process is
\begin{eqnarray}\label{stat_mdl_ols}
Y_j=z(t_j;\theta_0)+ \mathcal{E}_j & & \mbox{ for } j=1,\dots,n.
    \end{eqnarray}
    The errors $\mathcal{E}_j$ are assumed to be random variables satisfying the following assumptions:
    \begin{itemize}
        \item[(i)] the errors $\mathcal{E}_j$ have mean zero: $E[\mathcal{E}_j]=0$;
        \item[(ii)] the errors $\mathcal{E}_j$ have finite common variance: $\mbox{var}(\mathcal{E}_j)=\sigma_0^2<\infty$;
        \item[(iii)] the errors $\mathcal{E}_j$ are independent (i.e., $\mbox{cov}(\mathcal{E}_j,\mathcal{E}_i)=0$ whenever $j\neq i$)
        and identically distributed.
    \end{itemize}
Under these assumptions, we have that the mean of the observation
equals the model output: $E[Y_j]=z(t_j;\theta_0)$ and the variance
in the observations is constant in time:
$\mbox{var}(Y_j)=\sigma_0^2$.

\subsection{Ordinary least squares (OLS)}\label{ols}

We consider an ordinary least squares (OLS) formulation of a
generic parameter estimation or inverse problem for a vector
parameter ($\theta$)  dependent system
\begin{align}\label{vecsys}
\frac{dx}{dt}(t)&=g(t,x(t;\theta);\theta),\\
 x(t_0)&=x_0
\end{align}
with observation (or model output) process
\begin{equation}
z(t_j)=\mathcal{F}(x(\cdot),\theta), \quad j=1,\dots,n.
\end{equation}
In this context we consider a given vector of observations
$Y=(Y_1,\dots,Y_n)^T$, where each $Y_j$ is defined by equation
(\ref{stat_mdl_ols}), and the model output vector
$z(\theta)=(z(t_1;\theta),\dots,z(t_n;\theta))^T$ for a given
$\theta$.  The estimator $\theta_{OLS}=\theta_{OLS}^n$ is a random
variable that minimizes the Euclidian norm (in $\mathbb{R}^{n}$ )
square of $Y-z(\theta)$, i.e., $\theta_{OLS}^n$ minimizes
\begin{equation}\label{defnJ}
J(\theta|Y)\equiv\left| Y-z(\theta) \right|^2=
\left[Y-z(\theta)\right]^T\left[Y-z(\theta)\right]= \sum_{j=1}^{n}
\left[Y_j-z(t_j;\theta)\right]^2,
\end{equation}
which implies $\theta_{OLS}$ solves the gradient equation
\begin{equation}\label{optcondgls}
    \nabla_{\theta} (\left| Y-z(\theta)\right|^2) =0.
\end{equation}

Asymptotic theory can be used to describe the distribution of the
estimator $\theta_{OLS}$ \cite{banksdav,DG,sebwil}. Provided that
a number of regularity conditions as well as sampling conditions
are met (see \cite{sebwil} for details), it can be shown that,
asymptotically (i.e., as $n\rightarrow\infty$), $\theta_{OLS}$ is
approximately distributed according to a multivariate normal
distribution, i.e.,
\begin{equation}
\theta_{OLS}^n\sim\mathcal{N}_p\left(\theta_0,\Sigma_0^n\right),
\end{equation}
where $\Sigma_0^n=\sigma^2_0[n\Omega_0]^{-1}
\in\mathbb{R}^{p\times p}$ and
\begin{equation}
    \Omega_0=\lim_{n\rightarrow\infty}\frac{1}{n} \chi^{n}(\theta_0)^{T}\chi^n(\theta_0).
\end{equation}
We remark that the theory requires that this limit exists and that
the matrix $\Omega_0$ be non-singular. The matrix $\Sigma_0^n$ is
the $p\times p$ {\em covariance matrix}
$\mbox{cov}\left((\theta_{OLS}^n)_i,(\theta_{OLS}^n)_j\right)$,
and the $n\times p$ matrix $\chi(\theta_0)\equiv\chi^n(\theta_0)$
is called the {\em sensitivity matrix} of the system, and its
$j$th row is equal to $\nabla_{\theta}z(t_j;\theta_0)$.  More
precisely,
\begin{eqnarray} \label{chimtrxdef}
    \chi_{ji}^n(\theta_0)=\left.\frac{\partial z(t_j;\theta)}{\partial \theta_i} \right|_{\theta=\theta_0} && 1\leq j\leq n, \ 1\leq i\leq p.
\end{eqnarray}
For the motivating SEIRS model, the partial derivatives of the
state variable vector $x=(S,E,I,R)^T$ with respect to $\theta$ can
be readily calculated.  If $g=(g_1,g_2,g_3,g_4)^T$ denotes the
vector function whose entries are given by the expression on the
right sides of equations (\ref{seqn})--(\ref{xeqn}), then we can
write the seasonal SEIRS model in the general vector form
\eqref{vecsys}. The sensitivities $\partial x/\partial \theta$ are
calculated, for a given $\theta=\hat\theta_{OLS}$ (defined below),
by solving (see \cite{banksdav,ac08} and the references therein)
equation \eqref{vecsys} and then
    \begin{eqnarray}\label{seneqns}
        \frac{d}{dt}\frac{\partial x}{\partial \theta}&=&\frac{\partial g}{\partial x}\frac{\partial x}{\partial \theta}+\frac{\partial g}{\partial \theta},
    \end{eqnarray}
from $t=t_0$ to $t=t_n$.  In equation (\ref{seneqns}) the matrix
$\partial g/\partial x$ is $4\times4$, while the matrices
$\partial x/\partial \theta$ and $\partial g/\partial \theta$ are
$4\times p$.

The solution of equation (\ref{optcondgls}) obtained using a
realization $y=(y_1,\dots,y_n)^T$ of the observation process
$Y=(Y_1,\dots,Y_n)^T$ and denoted as the estimate
$\hat\theta_{OLS}=\hat\theta_{OLS}^n$, provides a realization of
the estimator $\theta_{OLS}$.  The estimate $\hat\theta_{OLS}$ is
used in the calculation of the sampling distribution for the
parameters.  The error variance $\sigma_0^2$ is approximated by
$\hat\sigma_{OLS}^2$, which is calculated as
\begin{equation}
    \hat\sigma^2_{OLS}=\frac{1}{n-p} \left| y-z(\hat\theta_{OLS})
    \right|^2.
\end{equation}
The covariance matrix $\Sigma_0^n$ is approximated by
$\hat\Sigma_{OLS}^n$, which is computed by
\begin{equation}\label{covmtrx}
    \hat\Sigma_{OLS}^n= \hat\sigma^2_{OLS} \left[\chi(\hat\theta_{OLS}^n)^T\chi(\hat\theta_{OLS}^n)\right]^{-1}.
\end{equation}
The approximation \cite{DG,sebwil} of the sampling distribution of
the estimator is
\begin{equation}
   \theta_{OLS}=\theta_{OLS}^n\sim \mathcal{N}_{p}(\theta_{0},\Sigma_{0}^n)\approx
    \mathcal{N}_{p}(\hat\theta_{OLS}^n,\hat\Sigma_{OLS}^n).
\end{equation}
The standard errors for $\hat\theta_{OLS}^n$ can be approximated
by taking the square roots of the diagonal elements of the
covariance matrix $\hat\Sigma_{OLS}^n$.  The standard errors are
used to quantify uncertainty in the estimation and are given by
\begin{equation}
SE_k(\hat\theta_{OLS}^n)=\sqrt{(\hat\Sigma_{OLS}^n)_{kk}}, \quad
k=1,\dots,p.
\end{equation}

\subsection{Fisher information matrix}
The matrix
\begin{equation}
F=F^n=\chi^n(\theta_0)^T\chi^n(\theta_0),
\end{equation}
is known as the Fisher information matrix \cite{banksded,cob80}.
Below, we use a linearization argument (similar to that employed
in the asymptotic distribution theory for OLS -- see Chapter 12 of
\cite{sebwil}) to give a heuristic derivation of an approximate
expression for the estimator $\theta_{OLS}$ in terms of $F$. This
derivation illustrates the role played by the Fisher information
matrix in the estimation of unknown parameters and uncertainty
propagation.


We observe that the gradient of $J(\theta|Y)$ as defined in
\eqref{defnJ} is given by
\begin{equation}
\nabla_{\theta}J(\theta|Y)=-2\chi^n(\theta)^T\left[Y-z(\theta)\right],
\end{equation}
because by equation (\ref{chimtrxdef}), we know that
$\nabla_{\theta}z(t_j;\theta)=  \chi^n(\theta_0)^T$. Moreover, the
Hessian of $J(\theta|Y)$ is
\begin{equation}
\nabla_{\theta}^2 J(\theta|Y)= 2\chi^n(\theta)^T\chi^n(\theta) -
G(\theta),
\end{equation}
where
    \[
    G(\theta)=2\sum_{j=1}^{n}\left[Y_j-z(t_j;\theta)\right]\nabla_{\theta}^2z(t_j;\theta).
    \]

For the next calculations we tacitly assume that
$\chi^n(\theta_0)^T\chi^n(\theta_0)$ is nonsingular and
$G(\theta_0)=0$. We consider a linearization of
$\nabla_{\theta}J(\theta|Y)$ around $\theta=\theta_0$, which is
given by
\begin{eqnarray}
\label{lnrztnjtehta}
\mathcal{L}(\theta)&=&-2\chi^n(\theta_0)^T\left[Y-z(\theta_0)\right]+
2\chi^n(\theta_0)^T\chi^n(\theta_0)(\theta-\theta_0).
\end{eqnarray}
The solution to $\mathcal{L}(\theta)=0$ is, to first order, the
minimizer $\theta_{OLS}^n$, and we thus have (see equation (2.15)
of \cite{sebwil})
\begin{equation}\label{newtonsolthetagls}
\theta_{OLS}^n\approx\theta_0+\left[\chi^n(\theta_0)^T\chi^n(\theta_0)\right]^{-1}\chi^n(\theta_0)^T\mathcal{E},
\end{equation} where $\mathcal{E}=(\mathcal{E}_1,\dots,\mathcal{E}_n)^T$, with
$\mathcal{E}_j=Y_j-z(t_j;\theta_0)$ for $j=1,\dots,n$.  The
propagation of uncertainty from the observation process to the
estimator is induced by $\mathcal{E}$ in equation
(\ref{newtonsolthetagls}).

It is clear from equation (\ref{newtonsolthetagls}) that if
$F^n=\chi^n(\theta_0)^T\chi^n(\theta_0)$ is nearly singular then
$\theta_{OLS}$ may be very sensitive to the observation error
$\mathcal{E}$.  Moreover, equation (\ref{covmtrx}) suggests that
near-singularity (or ill-conditioning \cite{golvan}) of $F^n$ may
also affect the approximation of the covariance matrix
$\hat\Sigma_{OLS}^n$, and consequently the calculation of standard
errors and confidence intervals for estimated parameters.

For some time it has been well understood (see
\cite{banksded,BEG,cob80,TC,yue08} and the references therein)
that the information content of measurements can be quantified by
the Fisher information matrix. Thus, efficient experiments can be
designed using the Fisher information matrix $F^n$. As noted in
 \cite{banksded}, the three most popular design strategies are: D-optimal design, c-optimal design, and
E-optimal design. These strategies involve the determinant, the
inverse, and maximum and minimum eigenvalues of $F^n$. Our
approach in this paper relies on properties of the sensitivity
matrix $\chi=\chi^n$ rather than $F^n$ as well as asymptotic
standard errors (which do depend on $F^n$) for parameters. In the
next section we address rank deficiency and the condition number
of the sensitivity matrix $\chi^n$.

\subsection{Singular value decomposition of the sensitivity matrix}

To motivate the role singular value decomposition plays in
uncertainty assessment, we consider another linearization that
relates the estimator $\theta_{OLS}$ to the singular values of the
rectangular sensitivity matrix $\chi$. (Hereafter we shall
suppress the superscripts denoting dependence on $n$ when no
confusion can occur.)

Suppose the model output $z(\theta)$ is well approximated by its
linear Taylor expansion around $\theta_0$, i.e.,
    \begin{equation} \label{linmodout}
        z(\theta)\approx z(\theta_0)+\chi(\theta_0)(\theta-\theta_0).
    \end{equation}
This first order Taylor expansion can be used to reduce
$Y-z(\theta)$ to an affine transformation of $\theta$, by using
equations (\ref{linmodout}) and (\ref{stat_mdl_ols}):
    \begin{equation}\label{affine}
        Y-z(\theta)=-\chi(\theta_0)(\theta-\theta_0)+\mathcal{E},
    \end{equation}
where $\chi(\theta_0)\in \mathbb{R}^{n\times p}$, $\theta-\theta_0
\in \mathbb{R}^{p}$, $\mathcal{E}$ is an $\mathbb{R}^{n}$-valued
random variable, and $n>p$.

The singular value decomposition (SVD) of the sensitivity matrix
$\chi(\theta_0)$ is denoted as
\begin{equation}\label{svdchi}
        \chi(\theta_0)=U\left[
                    \begin{array}{c}
                        \Lambda\\
                        \bf{0}
                    \end{array}
                    \right] V^T,
\end{equation}
where $U$ is an $n \times n$ orthogonal matrix, i.e.,
$U^TU=UU^T=I_n$, with $U_1$ containing the first $p$ columns of
$U$ and $U_2$ containing the last $n-p$ columns, $U=[U_1 \ U_2]$;
$\Lambda$ is a $p\times p$ diagonal matrix defined as
$\Lambda=\mbox{diag}(s_1,\dots,s_p)$, with $s_1\geq
s_2\geq\dots\geq s_p\geq0$; $\bf{0}$ denotes an $(n-p)\times p$
matrix of zeros; and $V$ denotes an orthogonal $p\times p$ matrix,
i.e., $V^TV=VV^T=I_p$ (more details about SVD can be found in
\cite{golvan} and references therein).

The Euclidean norm is invariant under orthogonal transformations.
In other words, for any vector $w\in\mathbb{R}^{n}$ we have that
$\left| w \right|^2=w^Tw=w^TIw=w^TUU^Tw=|U^Tw |^2$.  According to
\cite{golvan,nocwrig} this invariance of the Euclidean norm
implies
\begin{eqnarray} \label{svd2nor1}
\left| -\chi(\theta_0)(\theta-\theta_0)+\mathcal{E}  \right|^2&=& \left| U^T\left(-\chi(\theta_0)(\theta-\theta_0)+\mathcal{E}\right) \right|^{2} \\
        &=& \left| -\left[
                    \begin{array}{c}
                        \Lambda\\
                        {\bf 0}
                    \end{array}
                    \right] V^T (\theta-\theta_0) + \left[
                                    \begin{array}{c}
                                        U_1^T \\
                                        U_2^T
                                    \end{array}
                                    \right] \mathcal{E} \right|^{2} \\ \label{svdols}
        &=&  \left|-\Lambda V^T(\theta-\theta_0)+U_1^T\mathcal{E} \right|^{2} +  \left| U_2^T\mathcal{E}\right|^{2}.
\end{eqnarray}

The estimator $\theta_{OLS}$ minimizes $|Y-z(\theta) |^{2}$ and
according to equations \eqref{affine} and \eqref{svdols} can be
calculated by solving \(\left|-\Lambda
V^T(\theta-\theta_0)+U_1^T\mathcal{E} \right|^{2}=0\), for 
$\theta$ and thus obtaining
\begin{equation}\label{olsestsvd}
    \theta_{OLS}=\theta_0+V\Lambda^{-1}U_1^T\mathcal{E}= \theta_0+\sum_{i=1}^{p}\frac{1}{s_i}v_iu_i^T\mathcal{E},
\end{equation}
where $v_i\in\mathbb{R}^{p}$ and $u_i\in\mathbb{R}^{n}$ denote the
$i$th columns of $V$ and $U$, respectively (the matrix $V$ has
column partitioning $V=[v_1, \dots, v_p]\in\mathbb{R}^{p\times
p}$, while $U=[u_1,\dots,u_n]\in\mathbb{R}^{n\times n}$).

There is a similarity between equations (\ref{newtonsolthetagls})
and (\ref{olsestsvd}). Again, the randomness of the observation
process is additively propagated into the estimator. In equation
(\ref{olsestsvd}) we see that if $s_i\rightarrow0$, then the
estimator $\theta_{OLS}$ is particularly sensitive to
$\mathcal{E}$.

At this point we need a couple of definitions. The range of a
matrix $C\in\mathbb{R}^{n\times p}$ with column partitioning
$C=[c_1, \dots, c_p]$ is defined as the subspace spanned by its
columns, i.e.,
\begin{equation}
        \mathcal{R}(C)=\left\{\sum_{j=1}^{p}q_jc_j \in\mathbb{R}^{n}: q_j\in\mathbb{R}\right\}.
\end{equation}
The rank of a matrix $C\in\mathbb{R}^{n\times p}$ is equal to the
dimension of $\mathcal{R}(C)$:
    \begin{equation}\label{rankdef}
        \mbox{rank}(C)=\mbox{dim}(\mathcal{R}(C)).
    \end{equation}
If $\mbox{rank}(C)<\min\{n,p\}=p$ (because we are assuming there
are more observations than parameters, i.e., $n>p$) the matrix
$C\in\mathbb{R}^{n\times p}$ is said to be rank deficient.  On the
other hand, if $\mbox{rank}(C)=p$ we say the matrix
$C\in\mathbb{R}^{n\times p}$ has full (column) rank \cite{golvan}.

For a full rank sensitivity matrix
$\chi(\theta_0)\in\mathbb{R}^{n\times p}$ (assuming
$\mbox{rank}(\chi(\theta_0))=p$ and $s_1\geq s_2\geq\dots\geq
s_p>0$) its {\em condition number} $\kappa$ is defined as the
ratio of the largest to smallest singular value \cite{golvan}:
\begin{equation}
\kappa(\chi(\theta_0))=\frac{s_1}{s_p}.
\end{equation}

We note that if the matrix $\chi(\theta_0)$ has full rank and a
large condition number (a feature known as ill-conditioning
\cite{golvan}), then the Fisher information matrix
$F=\chi(\theta_0)^T\chi(\theta_0)$ inherits a large condition
number.
     Equation (\ref{svdchi}) implies the SVD of $\chi(\theta_0)^T\chi(\theta_0)$ is
    \begin{equation}
        \chi(\theta_0)^T\chi(\theta_0)=V\Lambda^2 V^T,
    \end{equation}
    and therefore
    \begin{equation}\label{cndnnbmrfshr}
        \kappa(\chi(\theta_0)^T\chi(\theta_0))=\frac{s_1^2}{s_p^2}=\left[\frac{s_1}{s_p}\right]^2=\kappa(\chi(\theta_0))^2.
    \end{equation}

As discussed in \cite{golvan}, if the columns of $\chi(\theta_0)$
are nearly dependent then $\kappa(\chi(\theta_0))$ is large.  In
other words, if $\kappa(\chi(\theta_0))$ is not large (the matrix
$\chi(\theta_0)$ is well-conditioned) then the columns of the
sensitivity matrix are not nearly dependent, suggesting one could
use the condition number of $\chi(\theta_0)$ as a criterium to
select parameter combinations.

In the next section we propose an algorithm for parameter
selection which is based on the rank and condition number of the
sensitivity matrix rather than the Fisher information matrix.

\section{Subset selection algorithm}\label{selalgo}

The identifiability analyses developed by Brun, et al.,
\cite{brun}, and Burth, et al., \cite{burth}, motivate the subset
selection algorithm introduced in this section. Both of these
approaches use submatrices of the Fisher information matrix in
their selection procedures. Burth, et al., implemented a
reduced-order estimation by determining which parameter axes lie
closest to the ill-conditioned directions of the Fisher
information matrix, and then by fixing the associated parameter
values at priori estimates throughout an iterative estimation
process.  The subset selection keeps the well-conditioned
parameters (those that can be estimated with little uncertainty
from given measurements) active in the optimization, subject to
having the corresponding Fisher information submatrix with a small
condition number.  Brun, et al., determine identifiability of
parameter combinations using the eigenvalues of submatrices that
result from excluding columns out of the Fisher information
matrix.  They quantify the near dependence of columns in the
sensitivity submatrix using the smallest eigenvalue of the Fisher
information submatrix.

We propose an algorithm that searches all possible parameter
combinations and selects some of them, based on two main criteria:
the full rank of the sensitivity matrix, and uncertainty
quantification as embodied in asymptotic standard errors.

Our approach is numerical and we illustrate its use with the SEIRS
model introduced earlier. To carry out the algorithm we require
prior knowledge of nominal variance and nominal parameter values.
We assume the observation error variance is $\sigma_0^2=500$, and
assume the following nominal parameter values for the SEIRS model:
\[
 \begin{array}{l}
            S_0= 2.78\times 10^{5} \mbox{(people)},\
            E_0=1.08\times 10^{-1} \mbox{(people)},\
            I_0=1.89\times 10^{-1} \mbox{(people)},\\
            N=1.00\times 10^{6} \mbox{(people)},\
            L=5.00 \mbox{(years)},\
            D=9.59\times 10^{-3} \mbox{(years)},\
            M=5.48\times 10^{-3} \mbox{(years)},\\
            P=75.00 \mbox{(years)},\
            \beta_0=375.00 \mbox{(years$^{-1}$)},\
            a_1=2.00\times 10^{-2},\
            b_1=-2.00\times 10^{-2}.
        \end{array}
    \]

Henceforth, we use the terms \lq\lq parameter combination" and
\lq\lq parameter vector" interchangeably.  Parameter vectors
$\theta\in\mathbb{R}^{p}$ will be considered for different fixed
values of $p$. When $p=11$ the parameter combination
    \begin{equation}\label{p11}
        \theta=(S_0,E_0,I_0,N,L,D,M,P,\beta_0,a_1,b_1)\in \mathbb{R}^{11},
    \end{equation}
with the nominal parameter values given above, produces a rank
deficient sensitivity matrix $\chi(\theta)$ for the SEIRS model.
For $p=3$ the only parameter combination considered here is that
of the transmission parameters, i.e.,
    \begin{equation}\label{p3}
        \theta=(\beta_0,a_1,b_1)\in \mathbb{R}^{3}.
    \end{equation}
Other parameter vectors for fixed values of $p=4,\dots,10$ are
considered in the following way. For each fixed $j=1,\dots,7$, and
therefore fixed $p=3+j$, we explore parameter vectors of the form
    \begin{equation}\label{qgnl}
        \theta=(\lambda_1, \lambda_2,\dots,\lambda_j,\beta_0,a_1,b_1)\in\mathbb{R}^{p},
    \end{equation}
    where for $k=1,\dots,j$,
    \[
        \lambda_k\in\{S_0,E_0,I_0, N,L,D,M,P\}=\mathcal{I},
    \]
such that no entries of $\theta$ in equation (\ref{qgnl}) are
repeated.

The set
    \begin{equation}
        \mathcal{S}_p=\{\theta=(\lambda_1, \lambda_2,\dots,\lambda_j,\beta_0,a_1,b_1)\in\mathbb{R}^{p}=\mathbb{R}^{3+j}|\ \lambda_k\in\mathcal{I},
        \ \lambda_k\neq \lambda_{m} \forall\ k,m =1,\dots,j\}
    \end{equation}
    collects the parameter vectors explored by a combinatorial search.

We define the set
    \begin{equation}\label{viableq}
            \Theta_p=\{\theta|\ \theta\in \mathcal{S}_p \subset \mathbb{R}^{p},\ \mbox{rank}(\chi(\theta))=p\},
    \end{equation}
where $\chi(\theta)$ denotes the $n\times p$ sensitivity matrix,
and its rank is defined by equation (\ref{rankdef}). By
construction, the elements of $\Theta_p$ are parameter vectors
that give sensitivity matrices with independent columns.

An important step in the selection procedure involves the
calculation of standard errors (uncertainty quantification) using
the asymptotic theory described in Section \ref{ols}. For every
$\theta\in \Theta_p$, we define a vector of {\em coefficients of
variation} $\nu(\theta)\in \mathbb{R}^{p}$ such that for each
$i=1,\dots,p$,
        \[
            \nu_i(\theta)=\frac{\sqrt{(\Sigma(\theta))_{ii}}}{\theta_i},
        \]
        and
        \[
            \Sigma(\theta)=\sigma_0^2\left[\chi(\theta)^T\chi(\theta)\right]^{-1}\in\mathbb{R}^{p\times p}.
        \]

In other words, the components of the vector $\nu(\theta)$ are the
ratios of each standard error for a parameter to the corresponding
nominal parameter value. These ratios are dimensionless numbers
that allow comparison even when parameters have substantially
different units and scales (e.g., $N$ is on the order of $10^{6}$,
while $a_1$ is on the order of $10^{-2}$).  Next, define
    \[
        \alpha(\theta)=\left| \nu(\theta) \right|.
    \]
We call $\alpha(\theta)$ the {\em parameter selection score}, and
remark that $\alpha(\theta)$ near zero indicates lower uncertainty
possibilities in the estimation while large values of
$\alpha(\theta)$ suggest that one could expect to find wide
uncertainty in at least some of the estimates.

In the optimization literature the term ``feasible'' usually
denotes a vector satisfying inequality or equality constraints.
Here we use this term in the context of identifiability: a
feasible parameter vector denotes a combination that can be
estimated from data with reasonable to little uncertainty. More
precisely, we say a given $\theta\in \Theta_p$ is a {\em feasible
parameter vector} if both $\alpha(\theta)$ and
$\kappa(\chi(\theta))$ are relatively small.

We summarize the steps of the algorithm as follows:
\begin{enumerate}
\item{\bf Combinatorial search.}  For a fixed $j=1,\dots,7$, and
hence fixed $p=3+j$, calculate the set
    \[
        \mathcal{S}_p=\{\theta=(\lambda_1, \lambda_2,\dots,\lambda_j,\beta_0,a_1,b_1)\in\mathbb{R}^{p}|\ \lambda_k\in\mathcal{I},
        \ \lambda_k\neq \lambda_{m} \forall\ k,m =1,\dots,j\}.
    \]
The set $\mathcal{S}_p$ collects all the parameter vectors
obtained from a combinatorial search.
\item {\bf Full rank test}.
Calculate the set of viable parameters $\Theta_p$ as
    \[
    \Theta_p=\{\theta|\ \theta\in \mathcal{S}_p \subset \mathbb{R}^{p},\ \mbox{rank}(\chi(\theta))=p\}.
    \]
\item {\bf Standard error test.}  For every $\theta\in \Theta_p$
calculate a vector of coefficients of variation $\nu(\theta)\in
\mathbb{R}^{p}$ by
    \[
            \nu_i(\theta)=\frac{\sqrt{(\Sigma(\theta))_{ii}}}{\theta_i},
        \]
        for $i=1,\dots,p$, and
        \(
            \Sigma(\theta)=\sigma_0^2\left[\chi(\theta)^T\chi(\theta)\right]^{-1}\in\mathbb{R}^{p\times p}.
        \)
Calculate the parameter selection score as
    \(
        \alpha(\theta)=\left|\nu(\theta) \right|.
    \)
    \end{enumerate}

To illustrate the algorithm we consider several values of $p$,
while using the MATLAB (The Mathworks, Inc.) routine \texttt{rank}
(this routine computes the number of singular values that are
greater than ``machine tolerance'').

\begin{figure}[h]
        \begin{center}
        \includegraphics[width=5in,height=4in]{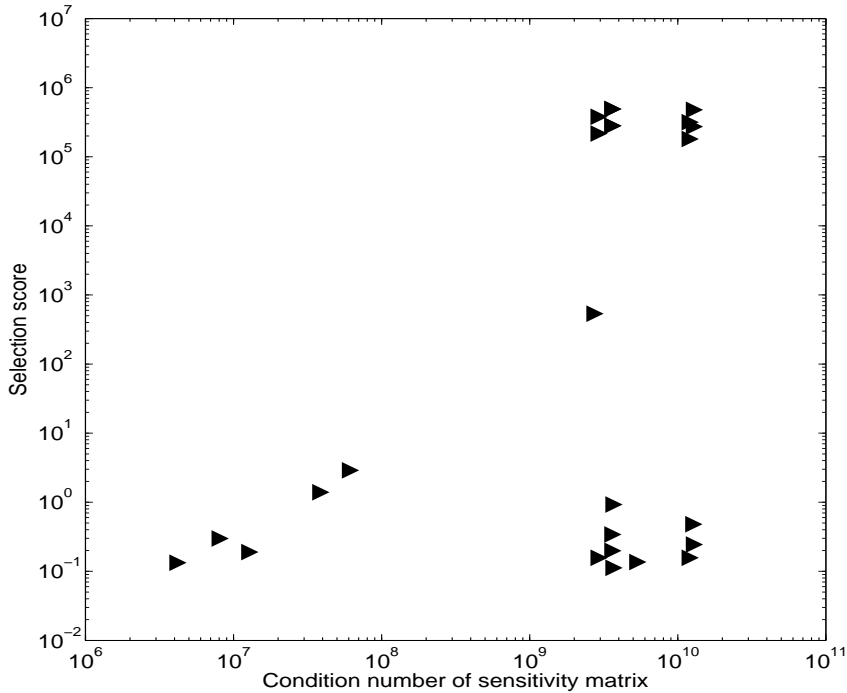}
        \end{center}
        \caption{Parameter selection score $\alpha(\theta)$ versus the condition number $\kappa(\chi(\theta))$ of the $n\times p$
        sensitivity matrix, for all parameter vectors $\theta\in \Theta_p$ with $p=5$.  Logarithmic scales
        are used on both axes.}
        \label{scorecondp5}
    \end{figure}

Results for $p=5$ (using the nominal parameter values) are
displayed in Figure \ref{scorecondp5} (on logarithmic scales),
where $\alpha(\theta)$ is depicted as a function of
$\kappa(\chi(\theta))$ for all $\theta\in \Theta_5$.  The pairs in
the lower-left corner of Figure \ref{scorecondp5} correspond to
feasible parameter vectors, because $\alpha(\theta)$ and
$\kappa(\chi(\theta))$ are here relatively small.

The subset selection algorithm was applied for $p=4,\dots,10$,
while using the nominal variance and parameter values.  We find
that there is not a single parameter combination with $p=10$ that
has a full rank sensitivity matrix.  For $p=9$, only three
parameter vectors pass the full rank test, and none of which can
be considered feasible.  We summarize the feasible parameter
vectors in Table \ref{tabfeaspar} for $p=4,\dots, 8$, where each
feasible $\theta \in \Theta_p$ is displayed along with
$\kappa(\chi(\theta))$ and $\alpha(\theta)$.  The cutoffs used to
select the parameter combinations in Table \ref{tabfeaspar} were
somewhat arbitrary but relative to the smallest values computed
for the two criteria (condition number and selection score) in
each example.

\begin{table}[h]
\caption{Feasible parameter vectors obtained while applying the
subset selection algorithm for $p=4,\dots,8$,  using nominal
values as listed earlier in the text. For each selected parameter
vector $\theta\in \Theta_p$ the condition number of the
sensitivity matrix $\kappa(\chi(\theta))$, and the selection score
$\alpha(\theta)$ are displayed.}
\begin{center}
\begin{tabular}{|c|c|c|} \hline\hline
Parameter vector $\theta$ & Condition number $\kappa(\chi(\theta))$ & Selection score $\alpha(\theta)$ \\ \hline \hline
$(L,\beta_0,a_1,b_1)$&2.047$\times 10^{5}$&5.019$\times 10^{-2}$\\ \hline
$(M,\beta_0,a_1,b_1)$&1.420$\times 10^{5}$&6.386$\times 10^{-2}$\\ \hline
$(P,\beta_0,a_1,b_1)$&3.176$\times 10^{5}$&7.044$\times 10^{-2}$\\ \hline \hline

$(L,D,\beta_0,a_1,b_1)$&4.034$\times 10^{6}$&1.332$\times 10^{-1}$\\ \hline
$(D,M,\beta_0,a_1,b_1)$&1.233$\times 10^{7}$&1.897$\times 10^{-1}$\\ \hline
$(D,P,\beta_0,a_1,b_1)$&7.781$\times 10^{6}$&2.987$\times 10^{-1}$\\ \hline \hline

$(N,L,D,\beta_0,a_1,b_1)$&1.829$\times 10^{10}$&1.670$\times 10^{-1}$\\ \hline
$(S_0,N,D,\beta_0,a_1,b_1)$&1.454$\times 10^{10}$&2.026$\times 10^{-1}$\\ \hline
$(S_0,L,D,\beta_0,a_1,b_1)$&1.828$\times 10^{10}$&2.375$\times 10^{-1}$\\ \hline
$(S_0,D,M,\beta_0,a_1,b_1)$&2.152$\times 10^{10}$&3.301$\times 10^{-1}$\\ \hline
$(S_0,D,P,\beta_0,a_1,b_1)$&1.828$\times 10^{10}$&4.832$\times 10^{-1}$\\ \hline
$(N,D,M,\beta_0,a_1,b_1)$&2.166$\times 10^{10}$&5.739$\times 10^{-1}$\\ \hline
$(N,D,P,\beta_0,a_1,b_1)$&1.829$\times 10^{10}$&9.658$\times 10^{-1}$\\ \hline \hline


$(N,L,D,M,\beta_0,a_1,b_1)$&2.166$\times 10^{10}$&5.960$\times 10^{0}$\\ \hline
$(S_0,L,D,M,\beta_0,a_1,b_1)$&2.167$\times 10^{10}$&5.970$\times 10^{0}$\\ \hline
$(N,D,M,P,\beta_0,a_1,b_1)$&2.166$\times 10^{10}$&1.153$\times 10^{1}$\\ \hline
$(S_0,D,M,P,\beta_0,a_1,b_1)$&2.167$\times 10^{10}$&1.159$\times 10^{1}$\\ \hline\hline

$(S_0,N,L,D,M,\beta_0,a_1,b_1)$&6.333$\times 10^{12}$&5.044$\times 10^{1}$\\ \hline
$(S_0,N,D,M,P,\beta_0,a_1,b_1)$&6.561$\times 10^{12}$&2.950$\times 10^{2}$\\ \hline
\end{tabular}
\end{center}
\label{tabfeaspar}
\end{table}


\section{Applications of the subset selection algorithm to synthetic and observed data sets}\label{appl}


The subset selection algorithm is illustrated first by solving
inverse problems from synthetic observations.  To construct a
synthetic data set we suppose a nominal parameter vector  and a
nominal error variance are equal to
 $\theta_0$ (true parameter vector) and $\sigma_0^2$ (true variance), respectively.  Random noise is then added
to the model output as follows:
\begin{equation}
    Y_j=z(t_j;\theta_0)+\sigma_0V_j,
\end{equation}
where $V_j$ is a standard normal random variable, i.e.,
$V_j\sim\mathcal{N}(0,1)$.  A realization $y_j$ of the observation
process $Y_j$, is calculated by drawing independent samples $v_j$
from the standard normal distribution so that
 \[
    \begin{array}{lr}
    y_j=z(t_j;\theta_0)+\sigma_0v_j   &\mbox{for $j=1,\dots,n$}.
    \end{array}
 \]

The OLS inverse problems were solved by implementing a subspace
trust region method (based on an interior-reflective Newton method
\cite{nocwrig}).  We used the MATLAB (The Mathworks, Inc.) routine
\texttt{lsqnonlin}.  For the purposes of this demonstration we
initialized every optimization routine at the nominal parameter
vector $\theta_0$.

The nominal error variance and nominal parameter values are those
given in the previous section. The parameter vectors estimated
from synthetic data are those appearing on top of each subtable in
Table \ref{syntdrslts}, for each value of $p$, where parameter
combinations are sorted in ascending order of their selection
score (from top to bottom).  In other words, all the parameter
vectors estimated from synthetic observations have reasonable
condition numbers and relatively small selection scores.  Five
inverse problems (for $p=8,7,6,5,4$) were solved from the same
realization of the observation process, to estimate the parameter
vectors
\[
\begin{array}{c}
\theta=(S_0,N,L,D,M,\beta_0,a_1,b_1), \\
\theta=(N,L,D,M,\beta_0,a_1,b_1), \\
\theta=(N,L,D,\beta_0,a_1,b_1),\\
\theta=(L,D,\beta_0,a_1,b_1),\\
\theta=(L,\beta_0,a_1,b_1).
\end{array}
\]
Results of these numerical experiments are summarized in Table
\ref{syntdrslts}.

\begin{figure}[h!]
\begin{center}
\includegraphics[width=4.5in,height=3.5in]{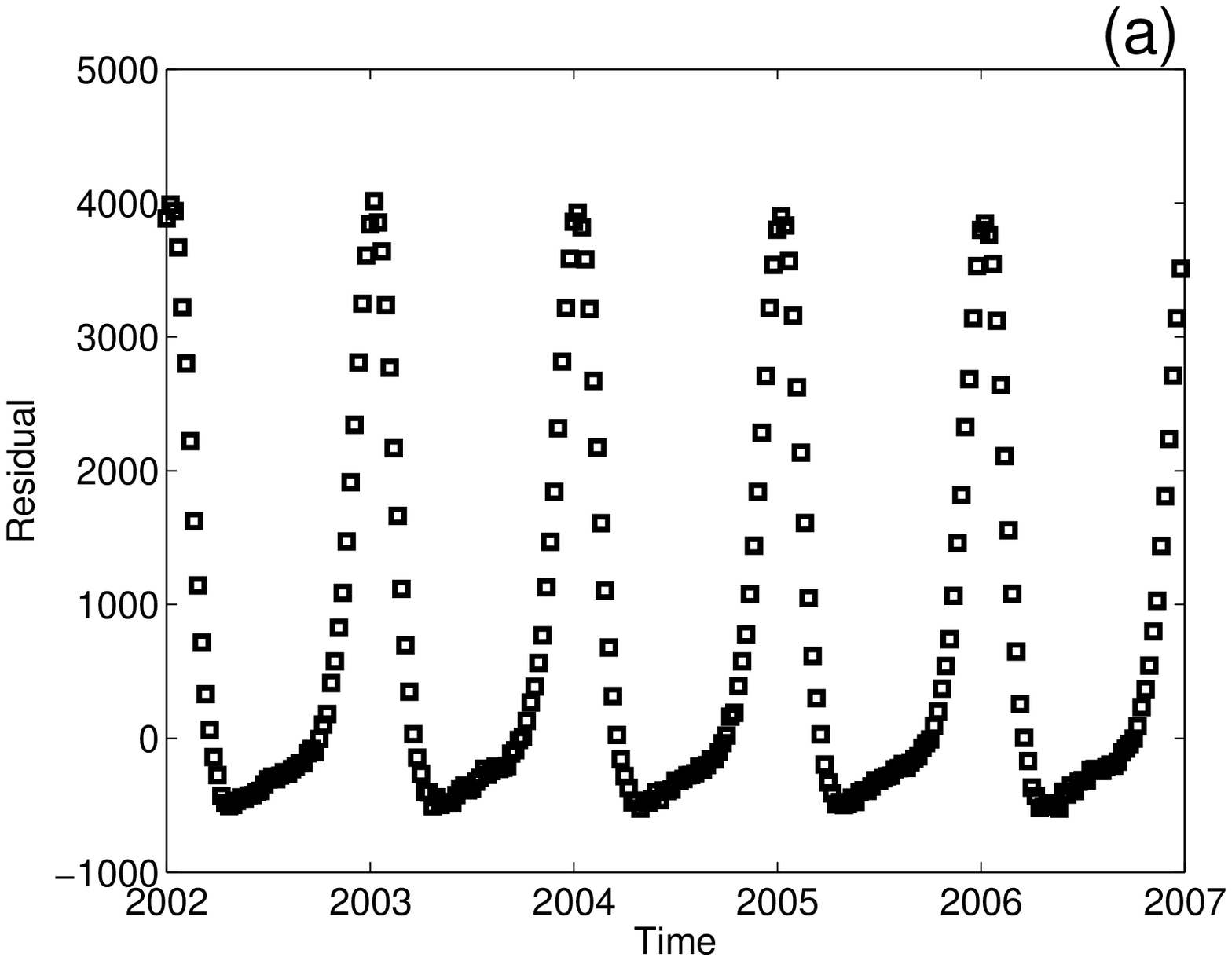}
\includegraphics[width=4.5in,height=3.5in]{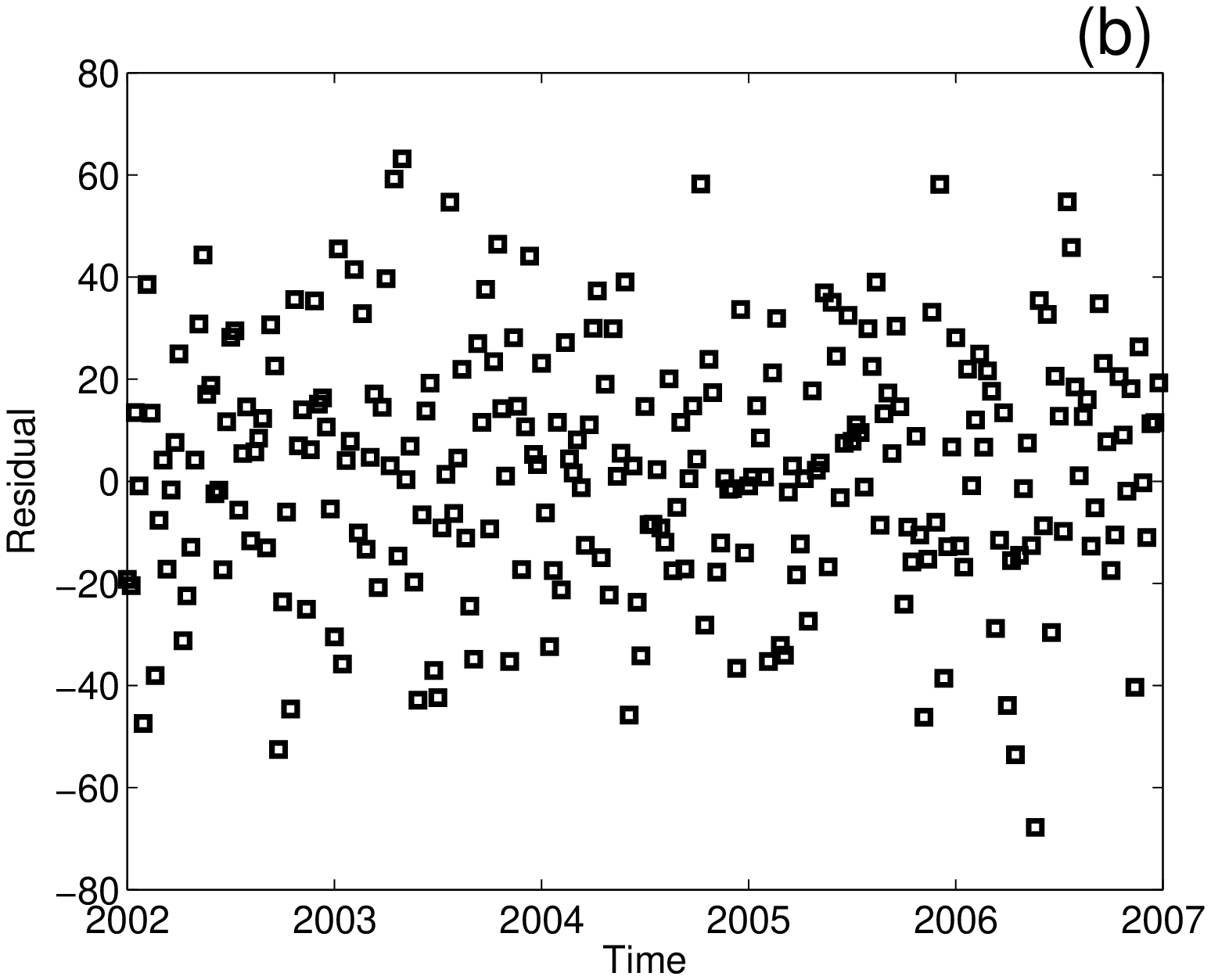}
\end{center}
\caption{Residual plots: $y_j-z(t_j;\hat \theta_{OLS})$, versus
time, $t_j$, for $j=1,\dots,n$.  Graph (a) displays residuals
obtained for $\theta=(S_0,N,L,D,M,\beta_0,a_1,b_1)$, while Graph
(b) depicts residuals for $\theta=(L,D,\beta_0,a_1,b_1)$.}
\label{ressynthd}
\end{figure}

\begin{table}[!h]
\caption{Results of solving five inverse problems from a single
synthetic data set generated as described in the text using
nominal values listed earlier. For each parameter combination we
display the estimate (Est.), the standard error (S.E.) and the
coefficient of variation (standard error divided by the estimate,
C.V. = S.E./Est.). For notational convenience we use here the
notation $e$ to denote exponentiation to the base 10; i.e., $2.8
e^5$ denotes $2.8\times 10^5$, etc.}
\begin{center}
\begin{tabular}{|c|c|c|c|c|c|c|c|c|} \hline
\multicolumn{9}{|c|}{Parameter vector $\theta=(S_0,N,L,D,M,\beta_0,a_1,b_1)$}\\
&$S_0$ &$N$&$L$&$D$&$M$&$\beta_0$&$a_1$&$b_1$ \\ \hline
Est.&2.8$e^{5}$&1.0$e^{6}$&5.0$e^{0}$&9.6$e^{-3}$&5.5$e^{-3}$&3.7$e^{2}$&2.0$e^{-2}$&-2.0$e^{-2}$\\
\hline
S.E.&1.5$e^{6}$&5.0$e^{6}$&4.5$e^{1}$&3.1$e^{-3}$&6.2$e^{-2}$&3.4$e^{3}$&7.7$e^{-2}$&8.4$e^{-2}$\\
\hline
C.V.&5.5$e^{0}$&5.0$e^{0}$&9.1$e^{0}$&3.2$e^{-1}$&1.1$e^{1}$&9.0$e^{0}$&3.8$e^{0}$&-4.2$e^{0}$\\
\hline\hline
\multicolumn{9}{|c|}{Parameter vector $\theta=(N,L,D,M,\beta_0,a_1,b_1)$}\\
Est.&
&1.0$e^{6}$&5.0$e^{0}$&9.6$e^{-3}$&5.5$e^{-3}$&3.7$e^{2}$&2.0$e^{-2}$&-2.0$e^{-2}$\\
\hline S.E.&
&2.7$e^{4}$&2.7$e^{0}$&2.5$e^{-3}$&2.2$e^{-2}$&5.9$e^{2}$&3.1$e^{-2}$&2.5$e^{-2}$\\
\hline C.V.&
&2.7$e^{-2}$&5.4$e^{-1}$&2.6$e^{-1}$&4.1$e^{0}$&1.6$e^{0}$&1.6$e^{0}$&-1.3$e^{0}$\\
\hline\hline
\multicolumn{9}{|c|}{Parameter vector $\theta=(N,L,D,\beta_0,a_1,b_1)$}\\
Est.&
&1.0$e^{6}$&5.0$e^{0}$&9.6$e^{-3}$&&3.8$e^{2}$&2.0$e^{-2}$&-2.0$e^{-2}$\\
\hline S.E.&
&2.7$e^{4}$&1.7$e^{-1}$&5.8$e^{-4}$&&1.5$e^{1}$&1.3$e^{-3}$&1.2$e^{-3}$\\
\hline C.V.&
&2.7$e^{-2}$&3.4$e^{-2}$&6.1$e^{-2}$&&3.9$e^{-2}$&6.3$e^{-2}$&-6.1$e^{-2}$\\
\hline\hline
\multicolumn{9}{|c|}{Parameter vector $\theta=(L,D,\beta_0,a_1,b_1)$}\\
Est.& &
&5.0$e^{0}$&9.6$e^{-3}$&&3.8$e^{2}$&2.0$e^{-2}$&-2.0$e^{-2}$\\
\hline S.E.& &
&7.4$e^{-2}$&5.8$e^{-4}$&&9.8$e^{0}$&1.2$e^{-3}$&1.2$e^{-3}$\\
\hline C.V.& &
&1.5$e^{-2}$&6.1$e^{-2}$&&2.6$e^{-2}$&6.2$e^{-2}$&-6.0$e^{-2}$\\
\hline\hline
\multicolumn{9}{|c|}{Parameter vector $\theta=(L,\beta_0,a_1,b_1)$}\\
Est.& & &5.0$e^{0}$& & &3.8$e^{2}$&2.0$e^{-2}$&-2.0$e^{-2}$\\
\hline S.E.& & &1.4$e^{-2}$& &
&2.6$e^{0}$&2.0$e^{-4}$&7.9$e^{-4}$\\ \hline C.V.& & &2.7$e^{-3}$&
&
&6.8$e^{-3}$&9.9$e^{-3}$&-4.0$e^{-2}$\\
\hline\hline
\end{tabular}
\end{center}
\label{syntdrslts}
\end{table}
We analyze the results using the coefficient of variation:
standard error (SE) divided by estimate (Est). For instance in
Table \ref{syntdrslts}, when
$\theta=(S_0,N,L,D,M,\beta_0,a_1,b_1)$ it is seen for $D$ that the
standard error is nearly one third of the estimate, suggesting
lower uncertainty.  For the other parameters $S_0$, $N$, $L$, $M$,
$\beta_0$, $a_1$, and $b_1$ the standard error can be nearly four
times (and up to eleven times) the estimate (for $b_1$ its SE is
$|4\times\mbox{Est}|$, because $b_1<0$).  This feature denotes
substantial uncertainty.  Figure \ref{ressynthd}(a) displays the
residual plot (see \cite{banksdav} for a discussion of the
effective use of residual plots) for this parameter combination:
$y_j-z(t_j;\hat \theta_{OLS})$ versus time $t_j$, where
$j=1,\dots,n$. The temporal pattern in the residuals together with
large standard errors suggest that estimation of this parameter
combination from observations (with a statistical error structure)
would be meaningless.

The residual plots for all the other parameter combinations in
Table \ref{syntdrslts} do not have temporal patterns. For the sake
of illustration we display in Figure \ref{ressynthd}(b) the
residuals versus time for $\theta=(L,D,\beta_0,a_1,b_1)$.

Improvements in uncertainty quantification are observed with the
removal of some key parameters. We think it is not just reducing
the number $p$ of parameters, but rather which parameters are to
be estimated that really counts.  The near dependence in the
columns of the sensitivity matrix $\chi$ reflects correlations
between parameter estimates which make a parameter combination
unsuitable for estimation. For instance, consider the removal of
$S_0$ from the estimation, and compare
$\theta=(S_0,N,L,D,M,\beta_0,a_1,b_1)$ with
$\theta=(N,L,D,M,\beta_0,a_1,b_1)$ in Table \ref{syntdrslts}. The
standard error for $N$ is seen to drop from 500\% to approximately
3\% of the estimate.  Another substantial improvement when
dropping $S_0$ is obtained for $L$, for which its standard error
reduces from being nine times the estimate to one half of its
value. Lower uncertainty improvements are also obtained for the
parameters $M$, $\beta_0$, $a_1$, and $b_1$.

The next numerical experiment considered here is the removal of
$S_0$ and $M$. We compare the results for
$\theta=(S_0,N,L,D,M,\beta_0,a_1,b_1)$ with those for
$\theta=(N,L,D,\beta_0,a_1,b_1)$, in Table \ref{syntdrslts}. There
are uncertainty improvements for all parameters.  The least (but
still substantial) improvement is for $D$, where its standard
error drops from being nearly 30\% to being just 6\% of the
estimate.  For the parameters $N$, $L$, $\beta_0$, $a_1$, and
$b_1$ an improvement of two orders of magnitude is seen.
Improvements in uncertainty are more pronounced after removing
$S_0$, $N$, and $M$: for this we compare
$\theta=(S_0,N,L,D,M,\beta_0,a_1,b_1)$ and
$\theta=(L,D,\beta_0,a_1,b_1)$ in Table \ref{syntdrslts}.

Undoubtedly, the best case scenario of uncertainty quantification
we obtained is that of estimating  $\theta=(L,\beta_0,a_1,b_1)$
from the same synthetic data set.  In Table \ref{syntdrslts}, it
is seen that the standard errors reduce to less than 1\% of the
estimates for $L$, $\beta_0$, and $a_1$, and to 4\% from nearly
400\% of the estimate for $b_1$.

As a final note in this section, we present results obtained from
solving the OLS problem while using observations of an
influenza-like-illness in France \cite{sentfranc}. Some of the
parameters were fixed to values suggested in
\cite{ciafactbook,chowmillvib,dusplo}:
\[
\begin{array}{l}
S_0= 1.56 \times 10^{7} \mbox{(people)},\ E_0=6.44
\mbox{(people)},\ I_0= 12.88 \mbox{(people)},\
N=6.40\times 10^{7}\mbox{(people)},\\
L=6.00\mbox{(years)},\ D=1.10 \times 10^{-2}\mbox{(years)},\ M=
5.50\times 10^{-3}\mbox{(years)},\ P= 80.87\mbox{(years)}.
\end{array}
    \]

The inverse problem was solved with $\theta=(\beta_0,a_1,b_1)$.
Simple inspection of the standard errors in Table \ref{ilitab8793}
does not seem to immediately suggest there is a poor fit (not
displayed here). Roughly speaking, the standard error is: 13\% of
the estimate for $\beta_0$; 30\% of the estimate for $a_1$; 45\%
of the estimate for $b_1$. These calculations give an indication
of wide uncertainty, but they are not as extreme as the results
for $\theta=(S_0,N,L,D,M,\beta_0,a_1,b_1)$ in Table
\ref{syntdrslts}. One can easily be misled by invalid uncertainty
quantification in the absence of residual analysis. Residual plots
(not displayed here) in this case have systematic patterns,
suggesting that either the statistical model (equation
(\ref{stat_mdl_ols})) may be incorrect, or more likely, the SEIRS
model fails to adequately describe the underlying process.

\begin{table}[h!]
\caption{Estimates from influenza-like-illness observations, where $ \theta=(\beta_0,a_1,b_1)$.  The coefficient
of variation is defined as the standard error divided by the estimate.}
\begin{center}
\begin{tabular}{|c|c|c|c|c|} \hline\hline
Parameter   &    Estimate    &    Standard error & Unit &Coefficient of variation\\ \hline
$\beta_0$&3.100$\times 10^{2}$&4.055$\times 10^{1}$&years$^{-1}$&1.308$\times10^{-1}$\\ \hline
$a_1$&1.539$\times 10^{-2}$&4.588$\times 10^{-3}$&1&2.981$\times10^{-1}$\\ \hline
$b_1$&-2.406$\times 10^{-2}$&1.090$\times 10^{-2}$&1&-4.530$\times10^{-1}$\\ \hline
\end{tabular}
\end{center}
\label{ilitab8793}
\end{table}

\section{Discussion}\label{discssn}

We have discussed a computational methodology for inverse problem
formulation in the context of parameter identifiability. Using an
OLS scheme based on a constant variance statistical model for the
observation process and a seasonal SEIRS epidemics model for
illustration, we have proposed a prior-analysis algorithm that we
believe might profitably precede efforts on parameter estimation
from data. The algorithm can be used if reasonable ranges for the
sought after parameters are either known a priori, or can be
assumed by the user much in the same way one must assume
reasonable ranges in inverse problem formulations and initiation
of algorithms for the resulting estimation procedures.

The subset selection \cite{miller90} algorithm given in Section
\ref{selalgo} is based on two main criteria for a fixed number of
parameters: (i) full rank of the sensitivity matrix; and (ii)
calculation of standard errors.  We proposed to first select
according to the sensitivity matrix rank, because those parameter
combinations for which $\chi$ has full rank will have a
non-singular Fisher information matrix $\chi^T\chi$, and its
inverse is used in the calculation of the standard errors (see
equation (\ref{covmtrx})).

The near dependence of the sensitivity matrix columns can be a
fingerprint of parameter correlations--a pertinent feature for
subset selection \cite{miller90}.  Capaldi, et al.,
\cite{capaldi08} determine identifiability of parameters in a
simple SIR model, and show how correlation between parameter
estimates can impede the estimation of other parameters and
parameter combinations, such as the basic reproductive number.
Moreover, Brun, et al., \cite{brun} explain that if the columns of
$\chi$ are nearly dependent, then changes in the model output due
to small changes in a single parameter can be compensated by
appropriate changes in other parameters.

We have presented illustrations of the how the removal of nearly
dependent columns of the sensitivity matrix can provide
substantial improvements in uncertainty quantification. This
feature involves more than just reducing the number $p$ of
parameters, it relates to excluding certain key parameters.  For
instance, if we assume a linear Taylor expansion of the model
output, the estimator $\theta_{OLS}\in\mathbb{R}^p$ is given by
equation (\ref{olsestsvd}), where the sensitivity matrix
$\chi(\theta_0)$ has singular values $s_1\geq\dots\geq s_{p-1}\geq
s_p>0$.  If $s_p \approx 0$ and $s_{p-1}>1$, then submatrices with
singular values $s_2\geq\dots\geq s_p>0$, and $s_1\geq\dots\geq
s_{p-1}$, have different conditioning when quantifying the
sensitivity of {\em reduced order} estimations that only involve
$p-1$ parameters. The condition number of the former submatrix is
$s_2/s_p$, which is large if $s_p \approx 0$, while for the latter
submatrix the condition number satisfies $1\leq s_1/s_{p-1}< s_1$,
because $s_{p-1}>1$.

In our numerical experiments, we calculate sensitivity matrices
$\chi(\theta)$ evaluated at different realizations of the
estimator $\theta=\hat\theta_{OLS}$.  When
$\theta=(S_0,N,L,D,M,\beta_0,a_1,b_1)$ the singular values of the
sensitivity matrix range from $4.7\times10^{6}$ to
$4.6\times10^{-6}$ while  for $\theta=(L,\beta_0,a_1,b_1)$ the
singular values of $\chi(\hat\theta_{OLS})$ range from
$1.9\times10^{6}$ to $9.3\times10^{0}$.

The smallest singular value changes from $4.6\times10^{-6}$ to
$9.3\times10^{0}$ while the largest remain on the order of
$10^{6}$. This improvement in conditioning is reflected in the the
standard error for $L$, $\beta_0$, and $a_1$, which reduces to
less than 1\% of the estimate, from nearly 900\% and 380\% (see
Table \ref{syntdrslts}).

Although in this paper we only discuss OLS, the selection
algorithm can be easily applied when using a generalized least
squares scheme \cite{banksdav}.  We also carried out numerical
experiments (for brevity not discussed here) involving use of
synthetic nonconstant variance data sets in GLS formulations, and
obtained results absolutely consistent with those of the OLS
formulation presented here (Section \ref{appl}).

%

%
\end{document}